\documentclass{article}[15pt]

\usepackage{amsmath,epsfig,color,calc,rotating}
\usepackage{wrapfig}

\begin{document}

\title{
Exploring Case-Control  Genetic Association Tests Using Phase Diagrams
\vspace{0.2in}
\author{
Wentian Li$^{1}$,
Young Ju  Suh$^{2}$, Yaning Yang$^{3}$  \\
{\small \sl 1. The Robert S. Boas Center for Genomics and Human Genetics,  The Feinstein Institute }\\
{\small \sl for Medical Research, North Shore LIJ Health System, 
 Manhasset, 350 Community Drive, NY 11030, USA.}\\
{\small \sl 2. Center for Genome Research, Samsung Biomedical Research Institute, Seoul 135-710, KOREA}\\
{\small \sl 3. Department of Statistics and Finance, University of Science and Technology of China,
 Anhui 230026, Hefei, CHINA}\\
}
\date{preprint version of {\sl Computational Biology and Chemistry}, 32(6):391-399 (2008)}
}
\maketitle  
\markboth{\sl Li, Suh, Yang) }{\sl Li, Suh, Yang}

\begin{center}
{\bf Abstract}
\end{center}

{\bf Backgrounds:}
By a new concept called ``phase diagram", we compare two 
commonly used genotype-based tests for case-control genetic analysis, 
one is a Cochran-Armitage trend test (CAT test at $x=0.5$, or CAT0.5) 
and another (called MAX2) is the maximization of two chi-square test results:
one from the two-by-two genotype count table
that combines the baseline homozygotes and heterozygotes, and
another from the table that combines heterozygotes with
risk homozygotes.  CAT0.5 is more suitable for multiplicative
disease models and MAX2 is better for dominant/recessive models.
{\bf Methods:}
We define the CAT0.5-MAX2 phase diagram on the disease model space 
such that regions where MAX2 is more powerful than CAT0.5 are
separated from regions where the CAT0.5 is more powerful,
and the task is to choose the appropriate parameterization to
make the separation possible. 
{\bf Results:}
We find that using the
difference of allele frequencies ($\delta_p$) and the difference
of Hardy-Weinberg disequilibrium coefficients ($\delta_\epsilon$)
can separate the two phases well, and the phase boundaries
are determined by the angle $tan^{-1}(\delta_p/\delta_\epsilon)$,
which is an improvement over the disease model selection 
using $\delta_\epsilon$ only. 
{\bf Conclusions:} We argue that phase diagrams similar to 
the one for CAT0.5-MAX2 have graphical appeals in understanding 
power performance of various tests, clarifying simulation schemes,
summarizing case-control datasets, and guessing the possible mode of inheritance.

\large

\section*{Introduction}

\indent

Comparing allele and genotype frequencies of a single marker 
between the patients and normal people (Sasieni, 1997; Lewis, 2002; 
Li, 2008) remains the core of a case-control genetic 
association analysis, prior to a haplotype analysis and bioinformatic analysis 
to determine the spatial extent and gene context of the signal.
The success of a genetic association study crucially depends on
study design (Amos, 2007), but the choice of test is also
somewhat important.
If we focus on genotype-based tests, i.e., each observed genotype
is a sample instead of two allele samples (Sasieni, 1997), there
are many possible types of tests to choose from. For example, the
Pearson's goodness-of-fit test on the 2-by-3 genotype count table,
Cochran-Armitage trend (CAT) tests with a single parameter value $x$
which determines the relative risk of heterozygote with
respect to the two homozygotes (Sasieni, 1997; Devlin and Roeder, 1999;
Slager and Schaid, 2001), tests that are maximized over two or more 
tests (Yamada et al., 2001; Freidlin et al., 2002; Tokuhiro et al., 2003;
Zheng et al., 2006a), and those that
stepwisely use a few tests sequentially (Zheng et al., 2008).
This quote from (Balding, 2006)  might capture the feeling about the 
current state: ``there is no generally accepted answer to the 
question of which single-(marker) test to use" in genetic case-control 
association analysis.  The widespread use of whole-genome association 
in the study of complex diseases nowadays (Wellcome Trust Case Control Consortium,
2007) adds an urgency in having a better understanding of this issue.

If one statistical test is always more powerful than another test,
the first test should of course be used all the time. However,
each test may have its own strength and weakness
depending on the true underlying model. For example, Cochran-Armitage 
trend as a family of tests is to test the equality of genotype frequencies
in the case and the control group, but as a test at a parameter $x$
is to test the null hypothesis $P_{case}(AA)+ x P_{case}(Aa)=
P_{control}(AA)+ x P_{control}(Aa)$
($AA$ is the risk genotype, and $P(AA)$ is its frequency; similarly
for heterozygous genotype $Aa$).  
The CAT test at $x=0$ is most powerful for detecting recessive 
disease genes, whereas it is more powerful at $x=0.5$ 
when Hardy-Weinberg equilibrium (HWE) holds for case 
and control groups. For a pair of tests, in the space of all possible 
disease models, the first test can be more powerful in 
some regions whereas the second test is more powerful in 
other regions. This simple picture is reminiscent of 
the phase diagram in physics (e.g., Gibbs, 1873; Lifshitz and Landau, 1980) 
where a phase can be solid, liquid or gas depending on the temperature, 
pressure, or other relevant quantities. In our case-control genetic test
example, the aim of phase diagram is to graphically depict regions
where specific test is more powerful than others.
Note that one should not confuse the usage of ``phase" here
with the meaning often used in genetics, i.e.,  of the
``parental origin" of an allele in a genotype
(e.g. Scheet and Stephens, 2006).

Although the principle of our phase diagram can be established
as discussed above,
there are several practical considerations. First, a disease model
is specified by many parameters, and one may wonder whether
the phase structure can be seen in a two-dimensional space.
Second, if the two phases are highly intermingled 
in one parameterization, we may ask if other parameterization schemes
are better suited for separating the two phases. Third, besides 
the definition of phases which are based on
statistical power, we can also define phases that are
based on $p$-value,
and the question is whether the two ways of defining phase diagrams are similar.
This paper uses a specific pair of tests to address these questions
and to show the utility of this approach.

The motivation of our study is to discuss the issue of in what circumstances,
one should use a particular test instead of others.
We first study the parameterization of phase diagram, then
construct two phase diagrams, one power-based and another $p$-value-based,
for a pair of commonly used tests. The phase boundary in both
versions of phase diagram will be determined. 
And we examine several whole-genome 
association datasets using the phase diagram. Several other
applications of the phase diagram are also discussed, including
comparison of two random simulations of disease models,
and graphical display of case-control data of many SNPs.

\section*{Methods}

\subsection*{Case-control difference of allele frequency and 
Hardy-Weinberg disequilibrium (HWD) coefficient given a 
disease model}

\indent

A single-locus biallelic disease model can be specified by 
four parameters:  three penetrances $f_i=$Prob(disease$|G_i$)
($i=0,1,2$) where $\{ G_i \}$ are the three genotypes $aa, aA, AA$, and
the population $A$-allele frequency $p$. Alternatively, one can use the
allele frequency $p$, two
relative genotype risks $\lambda_i= f_i/f_0$ ($i=1,2$),
and disease prevalence $K= f_0[ (1-p)^2 + 2p(1-p)\lambda_1 + p^2 \lambda_2]$
to specify the disease model. Denote $A$-allele frequency in
case and control group as $p_1$ and $p_0$, and HWD coefficient 
(Weir, 1996) in case and control group as $\epsilon_1$
and $\epsilon_0$; all four ($p_1, p_0, \epsilon_1, \epsilon_0$) can 
be derived from a given disease model,
under the assumption that HWE holds true for the general population that
contains both cases and controls (Wittke-Thompson et al., 2005). We 
use the difference of $A$-allele frequency ($\delta_p$) and the difference of 
HWD coefficients ($\delta_\epsilon$) in 
case and control group as the parameters for our phase diagram:
\begin{eqnarray}
\label{eq:delta-model}
\delta_p \equiv
p_1 - p_0 &=& \frac{f_0( p^2 \lambda_2 + p(1-p) \lambda_1)}{K}
- \frac{ p^2 (1-f_0\lambda_2)+p(1-p)(1-f_0\lambda_1)}{ 1-K}
 \nonumber \\
\delta_\epsilon \equiv
\epsilon_1 - \epsilon_0 &=& \frac{f_0^2p^2(1-p)^2 (\lambda_2-\lambda_1^2)}{K^2}
- \frac{f_0 p^2(1-p)^2 (2\lambda_1-1-\lambda_2-f_0\lambda_1^2+f_0\lambda_2)}
{(1-K)^2}
 \nonumber \\
& &
\end{eqnarray}
When $\lambda_1 > 1 $ and $\lambda_2 > 1$, $A$-allele is enriched in
the case group, and $\delta_p$ is always positive. On the other hand,
if $\lambda_1 < 1 $ and $\lambda_2 < 1$, $A$-allele is depleted in
the case group and $\delta_p$ is negative. To be consistent, we
call $A$ the risk allele when $\delta_p > 0$.  Whether $A$-allele is
a minor allele ($p < 0.5$) or a major allele ($p > 0.5$) does not
by itself affect the sign of $\delta_p$ and $\delta_\epsilon$.

\subsection*{Case-control difference of allele frequency and Hardy-Weinberg 
disequilibrium coefficient given a genotype count table}

\indent

Table 1(A) shows a case-control
genotype count table $\{ N_{ij} \}$ ($i=$1,0 for case, control,
and $j=$0,1,2 for genotypes with 0, 1, 2 copies of allele $A$)
with total $N=N_1+N_0$ samples.  The same table is parameterized 
in Table 1(B) using the estimated
$\{ \hat{p}_i \}$ and $\{ \hat{\epsilon}_i \}$ ($i=$1,0 for case, control).
The estimated $A$-allele frequency ($\hat{\delta}_p$) and the 
estimated difference of HWD coefficients ($ \hat{\delta}_\epsilon$) in
case and control group can be obtained from Table 1(A) as:
\begin{eqnarray}
\label{eq:delta-data}
\hat{\delta}_p &=& \hat{p}_1-\hat{p}_0 =  \frac{N_{11}/2+N_{12}}{N_1} -
\frac{N_{01}/2+ N_{02}}{N_0}
 \nonumber \\
\hat{\delta}_\epsilon &=& \hat{\epsilon}_1 - \hat{\epsilon}_0 =
\frac{N_{12}}{N_1} - \left( \frac{N_{11}/2+N_{12}}{N_1}  \right)^2
- \frac{N_{02}}{N_0} + \left( \frac{N_{01}/2+ N_{02}}{N_0} \right)^2
\end{eqnarray}
For notation simplicity, the hat ($\hat{}$) will be removed
later on, and whether the model-based or data-based usage is 
applied should be clear from the context.

Note that switching the $A$ and $a$ allele, or equivalently, switching
the first and the third column in Table 1, changes the sign of
$\delta_p$, whereas the sign of $\delta_\epsilon$ is unaffected.
Eq.(\ref{eq:delta-data}) also shows another advantage of using
the difference of two HWD coefficients:
if the disequilibrium is sensitive to typing errors, its effect
is minimized when the difference $\hat{\delta}_\epsilon$ is used.

\subsection*{Cochran-Armitage trend test and MAX2 test}

\indent

Cochran-Armitage trend (CAT) test at parameter $x$ assigns a 
score of 0, $x$, and 1 for genotypes $aa$, $aA$, and $AA$,
for log-risk relative to the baseline $aa$ genotype
(Sasieni, 1997; Devlin and Roeder, 1999; Slager and Schaid, 2001).
Sometimes, the name of Cochran-Armitage trend test 
is used to only refer to that at parameter $x=0.5$
(Balding, 2006), which we will call as CAT0.5. On the other
hand, CAT at parameter settings of $x=0$ and $x=1$ is equivalent
to assuming recessive and dominant models, which we will refer to
as CAT0 and CAT1.

With the genotype count in Table 1, the test statistic of 
CAT0.5, CAT0, and CAT1 can be derived (see, e.g., Sasieni, 1997).
We re-parameterize these test statistics using a new set of
parameters including $\delta_p$ and $\delta_\epsilon$:
\begin{eqnarray}
\label{eq:x2-catt}
X^2(CAT0.5) &=& \frac{ N_1N_0}{N} \cdot \frac{2\delta_p^2}
 { \overline{p}+\overline{p^2} - 2\overline{p}^2 + \overline{\epsilon}} 
 \nonumber \\
X^2(CAT0) &=& \frac{ N_1N_0}{N} \cdot
\frac{ (2\stackrel{=}{p}\delta_{p}+\delta_\epsilon)^2}
 { (\overline{p^2}+\overline{\epsilon})(1-\overline{p^2}-\overline{\epsilon})}
 \nonumber \\
X^2(CAT1) &=& \frac{ N_1N_0}{N} \cdot
\frac{ (2(1 - \stackrel{=}{p}) \delta_p - \delta_\epsilon)^2}
{(2\overline{p}-\overline{p^2}-\overline{\epsilon})
(1-2\overline{p}+\overline{p^2}+\overline{\epsilon}) }
\end{eqnarray}
where 
$\delta_p \equiv p_1-p_0$,
$\overline{p} \equiv (N_1/N)p_1 +(N_0/N)p_0$,
$\overline{p^2} \equiv (N_1/N)p_1^2 +(N_0/N)p_0^2$,
$\overline{\epsilon} \equiv (N_1/N)\epsilon_1 +(N_0/N)\epsilon_0$,
$\stackrel{=}{p} \equiv (p_1+p_0)/2$. 
CAT0.5 usually leads to very similar result to the allele-based
test. But since CAT0.5 is a genotype-based test, it does 
not have the problem of allele-based test for artificially doubling the 
sample size (Sasieni, 1997).

The MAX2 test is defined by the maximization of the CAT0 and CAT1
test statistics: $X^2(MAX2) \equiv \max( CAT0, CAT1)$. Although
MAX2 has been used in a few analyses (e.g., Yamada et al., 2001;
Tokuhiro et al., 2003), it did not have a formal name.  
The name of MAX2 used here is to distinguish it from
the MAX3 test ($\max( CAT0, CAT1, CAT0.5)$)
proposed in (Freidlin et al., 2002).

Since MAX2 involves a multiple testing whereas CAT0.5 does not,
its $p$-values are calculated differently.  For CAT0.5, the 
$p$-value is simply derived by the $\chi^2$ distribution
with 1 degree of freedom. For MAX2, Dunn-\u{S}id\'{a}k multiple testing correction
(Ury, 1976) is exact if CAT0 and CAT1 test statistics were independent:
\begin{eqnarray}
\label{eq:dunn}
p_{MAX2} &=& P_0( X^2(MAX2) > T) = 1- P_0( X^2(MAX2) < T) = 1- P_0(X^2 < T)^2
\nonumber \\
& = & 1- (1-p_{X2})^2 = 2p_{X2} - p_{X2}^2
\end{eqnarray}
where $P_0$ is the null distribution, $p_{X2}$ is the $p$-value 
for a $\chi^2$-distributed test statistic. 
In R code
({\sl http://www.r-project.org/}), the command for calculating $p$-value for
MAX2 is 
{\tt 1-pchisq(MAX2, df=1){\char94}2}.

For dataset generated by a known disease model, CAT test statistics
follow chi-square distributions with non-central parameters. The
non-central parameters for CAT0.5, CAT0 and CAT1 are given in
Eq.(\ref{eq:x2-catt}), only that $\delta_p$, $\delta_\epsilon$,
and other parameters are determined by the disease model, not from 
the data. Alternatively, power can be determined empirically by simulation.

\subsection*{Simulation}

\indent

For empirical power calculation, we sampled 
$N_r$ (=5000)  replicates of genotypes for 
$N_{case}$ (=500) cases and 
$N_{control}$ (=500) controls, given a disease model. 
In Fig.\ref{fig2},  $N_m$ (=10000)  disease models 
were generated randomly.  The relative genotype
risks $\lambda_1$ and $\lambda_2$ are randomly selected from a
range (e.g., (0.5-2)), and the population $A$-allele frequency is
randomly selected (e.g., from (0.1-0.9)). The type I error is set at 0.05 and 
we have determined the test statistic threshold 
for MAX2 either by permutation or by Dunn-\u{S}id\'{a}k formula.
Due to the consistency between the two approaches, the type I
error is controlled mostly by using the Dunn-\u{S}id\'{a}k formula
in Eq.(\ref{eq:dunn}).  
The empirical power of CAT0.5 or MAX2 
at the given type I error is determined by 
the proportion of replicates that exceeds the threshold. 

For the phase diagram in Fig.\ref{fig4}  under the null model
(i.e., same allele and genotype frequency according to the HWE), 
$N_r$ (=2000) replicates of genotype were generated for  $N_{case}$ 
(=500) cases and $N_{control}$  (=500) controls.

\section*{Results}

\subsection*{Phase diagrams based on power given the disease model}

\indent

\begin{figure}[t]
  \begin{turn}{-90}
   \epsfig{file=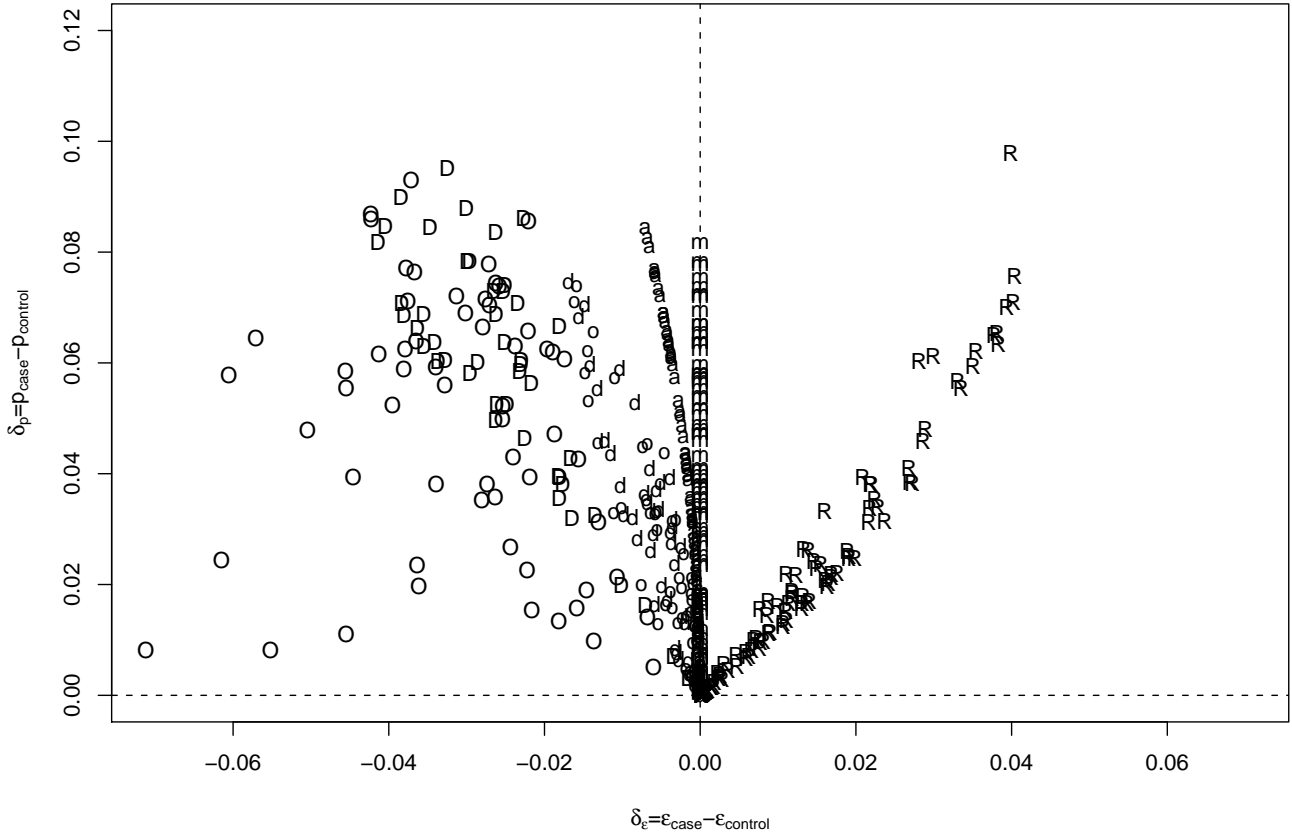, width=11cm}
  \end{turn}
\caption{
\label{fig1}
Locations in the phase diagram parameterized by $x=\delta_\epsilon$ and
$y=\delta_p$ for some commonly encountered disease models, recessive (R/r),
dominant (D/d), additive (A/a), multiplicative (M/m) and over-dominant (O/o).
The upper case is used when the power of MAX2 is larger than the power
of CAT0.5, and lower case for the opposite. The model parameters are sampled
from these ranges, population $A$-allele frequency is randomly sampled
from (0-0.5), phenocopy rate $f_0$ from ($10^{-6}$-$10^{-3}$), with the exception
of over-dominant models, $\lambda_2$
from (1.001-2) and $\lambda_1$ is derived from $\lambda_2$
($\lambda_1=\lambda_2$, 1, $(\lambda_2+1)/2$, $\sqrt{\lambda_2}$ for
for dominant, recessive, additive models, and multiplicative models).
For over-dominant models, $\lambda_1$ is randomly chosen from (1.001-2),
and $\lambda_2$ from (1.001, $\lambda_1$).
}
\end{figure}

In the $\delta_\epsilon$-$\delta_p$ space, known types of disease models
such as dominant, recessive, additive, multiplicative, and
over-dominant models fall in different regions of the plane,
after requiring the risk allele $A$ to have higher frequency in
cases than in controls, as can be seen from Fig.\ref{fig1}. 
For example, recessive models reside in the first quadrant,
additive, dominant, over-dominant models are in the second
quadrant, and the multiplicative models sit along the $y$-axis.

\begin{figure}[t]
  \begin{turn}{-90}
   \epsfig{file=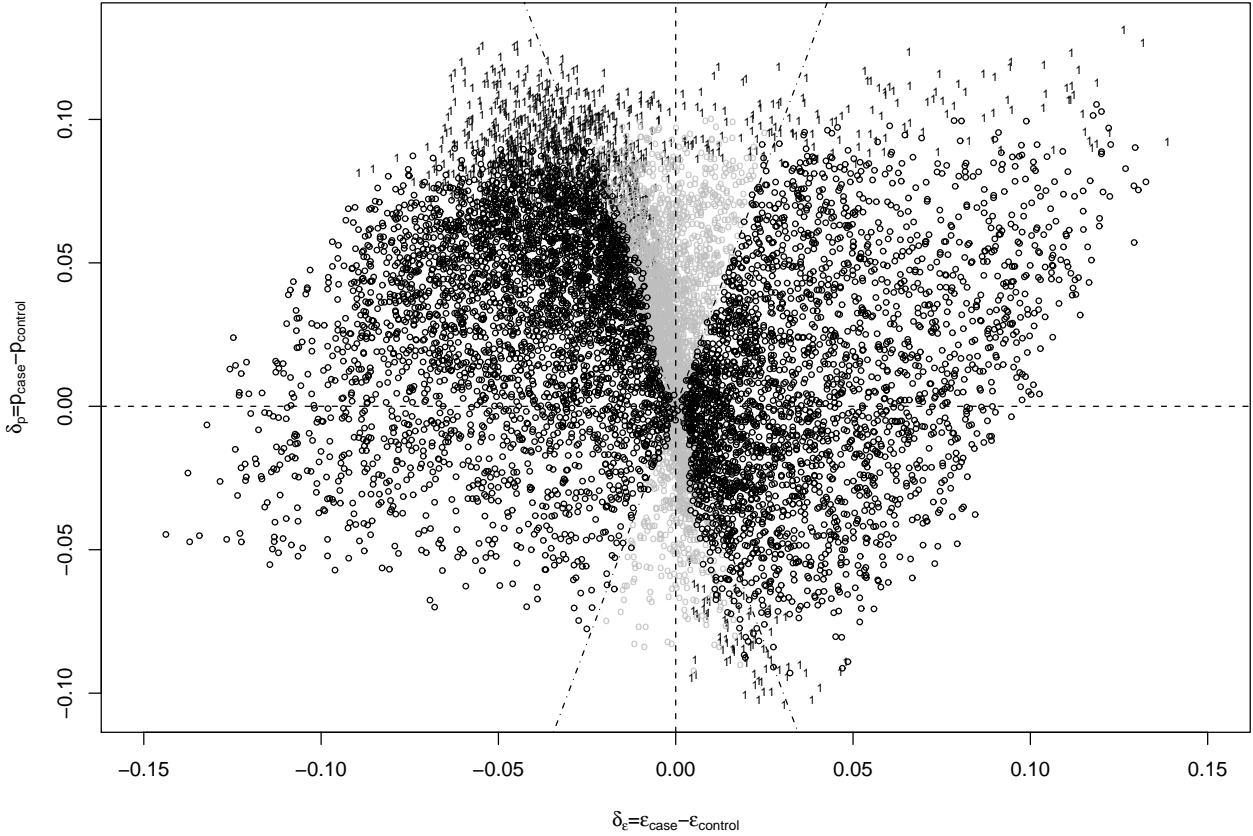, width=11cm}
  \end{turn}
\caption{
\label{fig2}
The CAT0.5-MAX2 phase digram, 
with $\delta_\epsilon=\epsilon_1-\epsilon_0$ (difference of two HWD
coefficients in case and control group) as the $x$-axis, and
$\delta_p=p_1-p_0$ (difference of allele frequencies in case and control
group) as the $y$-axis, determined by comparing the empirical power 
of CAT0.5 and MAX2 tests.  More than 10000 disease models are 
randomly sampled, each is used to randomly generate 5000 replicates of 
500 cases and 500 controls. The empirical power of CAT0.5 and MAX2 
based on these 5000 replicates are compared when the type I error is 
controlled at 0.05. A black circle is drawn when the  empirical power of
MAX2 is larger than that of CAT0.5, and a gray circle when CAT0.5 is more powerful 
than MAX2. If the empirical powers of both tests are larger than 0.99, 
a symbol ``1" is marked. The two dashed lines form angles of 73.125$^\circ$ 
and 106.875$^\circ$ ($13\pi/32$ and $19\pi/32$) with the $x$-axis, or, 
$\pm 16.875^\circ$ ($3\pi/32$) with the $y$-axis.
}
\end{figure}

We pick CAT0.5 and MAX2 as the two tests to compare for
the following reasons.  First, allele-based test is still
one of the most commonly used tests in case-control genetic
analysis, and we would like to choose a similar genotype-based 
test. Second, we want to choose a test that 
is robust against disease model mis-specification. These two
considerations lead to CAT0.5 and MAX2. Fig.\ref{fig2} 
shows the power-based CAT0.5-MAX2 phase diagram,
where empirically obtained statistical power of CAT0.5 and MAX2 
are compared while controlling the type I error, in the 
space parameterized by $\delta_\epsilon$ and $\delta_p$.

The region in Fig.\ref{fig2} where power(MAX2) $>$ power(CAT0.5),
or phase 1, covers most of the model space. On the other hand,
region where power(CAT0.5) $>$ power(MAX2), or phase 2, is
limited to a narrow angle around the $y$-axis. For regions far
away from the $x$-axis, both tests lead to close to 100\% power
(the symbol ``1" is used to mark the points when both 
power(CAT0.5) and power(MAX2) are larger than 0.99).
The phase boundary in the first (and the third) quadrant can be 
roughly approximated by the line 
$\theta= tan^{-1} (\delta_p/\delta_\epsilon)= 73.125^\circ$
($13\pi/32$).  The phase boundary in the second (and the fourth) quadrant 
is not sharp with some degree of overlap between the two phases.
However, the line $\theta= tan^{-1} (\delta_p/\delta_\epsilon)= 106.875^\circ$
($19\pi/32$) seems to provide a 
reasonable boundary to phase 2 points.

The phase structure presented in Fig.\ref{fig2} is
consistent with our current knowledge that allele-based
test and CAT0.5 are most powerful for multiplicative models.
In fact, it can be shown that non-central parameter in
the $\chi^2$ distribution for the allele-based test is
strictly larger than that of either CAT0 or CAT1  (i.e.,
either one of two ways to combine heterozygote counts with 
the homozygote counts), if (1) $\epsilon_1=\epsilon_0=0$ 
and (2) $N_1=N_0$ (Suh and Li, 2007).  However, it was somewhat surprising 
that the two phases in Fig.\ref{fig2} are not 
separated by vertical lines.  The two-phase genetic model
selection method proposed in (Zheng and Ng, 2008)  attempts
to infer the underlying disease model by $\delta_\epsilon$
value alone. Here we show that the model selection
could be more accurate if both $\delta_\epsilon$ and $\delta_p$
parameters are considered.

\begin{figure}[t]
  \begin{turn}{-90}
   \epsfig{file=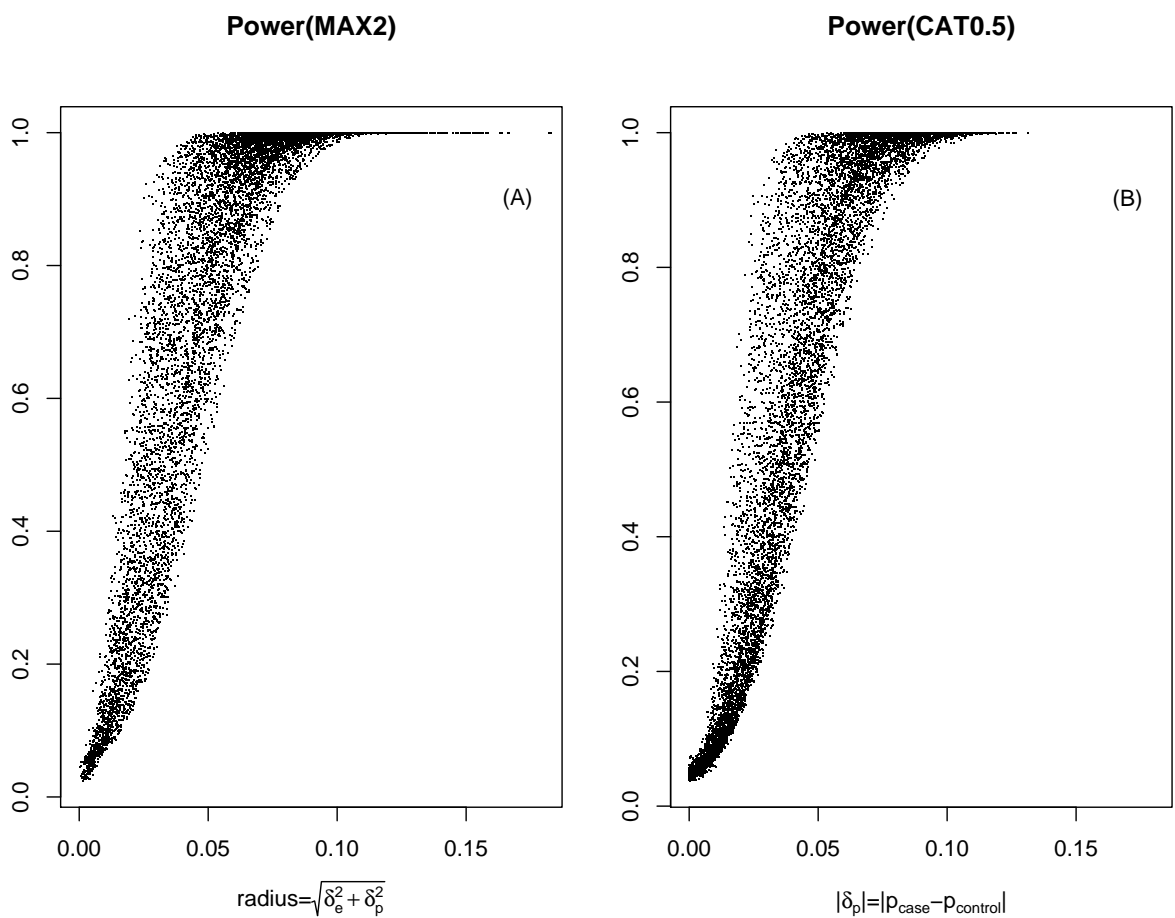, width=12cm}
  \end{turn}
\caption{
\label{fig3}
(A) The statistical power of MAX2 as a function of the radius
in the $\delta_\epsilon-\delta_p$ plane,
$\sqrt{ \delta_\epsilon^2+\delta_p^2}$.
(B) The statistical power of CAT0.5 as a function of vertical
distance to the $x$-axis, $|\delta_p|$.
}
\end{figure}

To have a better understanding of the phase structure, in Fig.\ref{fig3}, 
we plot the power of CAT0.5 as a function of $ |\delta_p| $ as 
well as power of MAX2 as a function of radius 
$\sqrt{ \delta_p^2 + \delta_\epsilon^2}$. As expected, the power 
of CAT0.5 increases as the allele frequency difference is larger, 
because CAT0.5 is very close to the allele-based test which is 
designed to detect allele frequency differences.  On the other 
hand, it was unexpected that the power of MAX2 increases with the radius,  
and the increase follows a similar way as power of CAT0.5 increases with
$|\delta_p|$. If we draw the two plots in one, the two more or
less overlap. In a crude approximation, suppose power(CAT0.5)
vs. $|\delta_p|$ and power(MAX2) vs. $\sqrt{ \delta_p^2 + \delta_\epsilon^2}$
follow the same functional form, then the power of MAX2 should
be larger than that of CAT0.5 over the whole plane except
the $y$-axis, because radius is always larger than the $y$ value
except for the $y$-axis. Although this is only an approximation, 
Fig.\ref{fig3} helps us to understand why the robust
MAX2 test is more likely to be more powerful than the CAT0.5 test
in the space of all possible models.

\subsection*{Phase diagrams based on $p$-values given the genotype data}

\indent

Power analysis is always discussed in a theoretical context because
it requires our knowing of the true disease model. In reality,
the disease model is not supposed to be known.  Towards a data-driven 
concept of phase diagram, we define the following phase
diagram based on $p$-values. 

As can be seen from Eq.(\ref{eq:x2-catt}), the relative magnitude of 
CAT0.5 and MAX2 test statistics depends on the estimated value of four parameters
$p_1, p_0, \epsilon_1, \epsilon_0$, as the common factor 
$N_1 N_0/N$ is canceled out.
If we assume that averaged quantities such as
$\overline{p}$ and $\overline{\epsilon}$ do not vary dramatically
with different $p_1, p_0, \epsilon_1, \epsilon_0$ values,
then the relative magnitude of CAT0.5 and MAX2 test statistics are 
mainly determined by the parameters $\delta_p$ and $\delta_\epsilon$.

Because of the multiple testing in MAX2, even if a MAX2 test
statistic is larger than that of CAT0.5, the MAX2 $p$-value
may not be smaller than the $p$-value by CAT0.5.
It should be noted that the assumption to apply Dunn-\u{S}id\'{a}k 
formula in Eq.(\ref{eq:dunn}), i.e., the independence between 
the two terms to be maximized, does not hold exactly.
However, the correlation between CAT0 and CAT1 is actually small. 
It was shown in (Freidlin et al., 2002; Zheng et al., 2006a)  that the correlation
between CAT0 and CAT1 test statistics under the null
hypothesis ($p_1=p_0$), if HWE is true, 
is $\sqrt{ p_1 (1-p_1)/[ (1+p_1)(2-p_1)] }$. This correlation is 
at most 1/3 which is reached when $p_1=p_0=0.5$. 
Simulation shows that
Dunn-\u{S}id\'{a}k formula leads to an accurate estimation of $p_{MAX2}$,
despite a weak violation of the assumption.
This is in a contrast with MAX3, where the correlation
between CAT0.5 and either CAT1 or CAT0 is too strong to apply the
Dunn-\u{S}id\'{a}k formula (Gonz\'{a}lez et al., 2008; Li et al., 2008).

\begin{figure}[t]
  \begin{turn}{-90}
   \epsfig{file=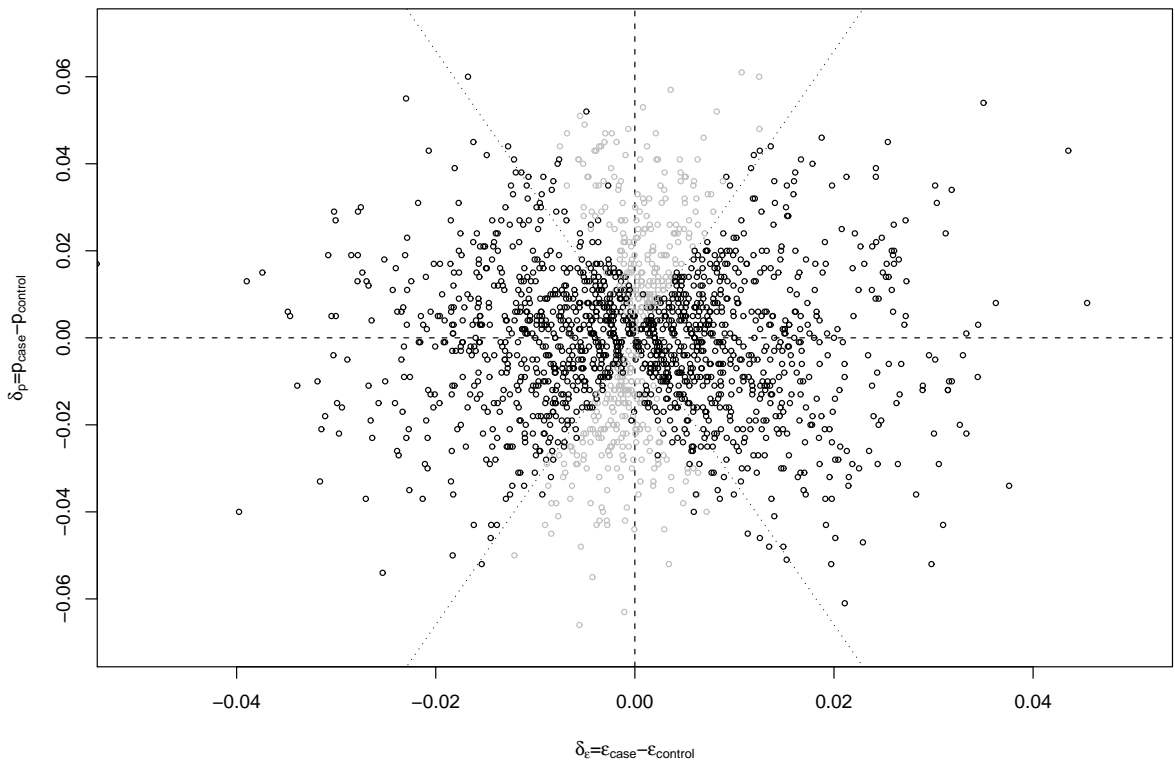, width=10cm}
  \end{turn}
\caption{
\label{fig4}
The CAT0.5-MAX2  phase diagram based on $p$-value  of two tests.
2000 genotype count tables are simulated using the null model
with 500 cases and 500 controls, and
$A$-allele frequency being samples from the range of ($10^{-5}$-0.5).
If the $p$-value by MAX2 is smaller (larger) than that by CAT0.5,
a black (grey) circle is drawn. The two dashed lines are the same
as those in Fig.\ref{fig2}, with angles of 73.125$^\circ$ and 106.875$^\circ$ 
($13\pi/32$ and $19\pi/32$) with the $x$-axis. 
}
\end{figure}

Fig.\ref{fig4} shows the phase diagram based on comparing $p$-values
of CAT0.5 and MAX2 when genotype count tables are randomly sampled
from the null model. The MAX2 (CAT0.5) test is preferred over CAT0.5 
(MAX2) in phase 1 (2) because it has a smaller $p$-value (black (grey) dots).
When we draw the same straight lines as Fig.\ref{fig2} with 73.125$^\circ$ and
106.875$^\circ$,  a similar observation can be made that the phase
separation is much better in the first (and the third) quadrant,
and a low level of overlap occurs in the second (in the fourth) quadrant.
We conclude that the CAT0.5-MAX2 phase diagram based on $p$-values is very
similar to the CAT0.5-MAX2 phase diagram based on power. Because
the ranges of $x$ and $y$ in Fig.\ref{fig4} are smaller than
those in Fig.\ref{fig2}, another understanding of $p$-value-based
phase diagram is that it is the ``extension" of power-based
phase diagram towards the origin.

\subsection*{Illustration of the CAT0.5-MAX2 phase digram by
results of several genome-wide association studies}

\indent

Recently, the largest scale whole-genome association study was
carried out by Wellcome Trust Case Control Consortium on several common
complex diseases (Wellcome Trust Case Control Consortium, 2007).
We take the genotype count for SNPs that showed the strongest,
and for many of them, validated association signal using extra
samples, to illustrate the use of phase diagram. The association signal 
in (Wellcome Trust Case Control Consortium, 2007) s mainly ranked by the CAT0.5 test.

Table 2(A) lists the raw genotype count data of 11 SNPs that are
associated with one of these diseases: ankylosing spondylitis 
(Wellcome Trust Case Control Consortium \&  Australo-Anglo-American
Spondylitis Consortium, 2007), type 1 diabetes (Todd et al., 2007), 
Crohn's disease (Parkes et al., 2007),  and type 2 diabetes 
(Zeggini et al., 2007).  Table 2(B) shows the estimated case or 
control minor allele frequency ($p_1, p_0$) and their difference 
($\delta_p$), case or control HWD coefficients ($\epsilon_1, \epsilon_0$) 
and their difference ($\delta_\epsilon$), the angle $\theta$ in 
the $\delta_p-\delta_\epsilon$ phase diagram with respect to either $x$ or $y$ axis, and
CAT test statistics at $x=0.5, 0, 1$.

There are several observations made from Table 2(B). As expected,
the estimated HWD coefficient in case group $\epsilon_1$
is usually larger in magnitude than that in control group $\epsilon_0$.
However, the largest observed $\epsilon_1$ is only around 0.01. On the other
hand, allele frequency difference is large (0.03-0.07) due to the
fact that these SNPs are selected by significant CAT0.5 test.
A consequence of the two facts is that the angle with respect to the $y$-axis
in Table 2(B) tends to be small, with the exception of SNP rs27044
which is associated with ankylosing spondylitis ($\theta-90^\circ= 16.4^\circ$). 

The closeness to the $y$ axis of these SNPs on the phase diagram 
should indicate, on average,  CAT0.5 test to be better than MAX2 test. Indeed,
the $X^2$(CAT0.5) test statistics are larger than $X^2$(MAX2)
except for two SNPs: rs27044 and rs2542151. It is not surprising to see
rs27044 in the exception list as it has the largest angle with respect
to the $y$-axis, and the negative sign of $\delta_\epsilon$
indicates that the disease model for rs27044 is more likely dominant. 
For rs27044, MAX2 test leads to a more significant result 
($p$-value = 1.8 $\times 10^{-7}$) than the CAT0.5 test
($p$-value= 10$^{-6}$). For rs2542151, its position in the phase
diagram forms a smaller angle with the $y$-axis than the rs27044
SNP (8.5$^\circ$ vs. 16.4$^\circ$),
but the angle is still large enough such that MAX2 leads to a more significant 
test ($p$-value =$3.7 \times 10^{-14}$) than CAT0.5 test
($p$-value = $1.9 \times 10^{-13}$).

\section*{Discussion and conclusions}

\indent

Our choice of $\delta_p$ is to capture the linear or first-order term
from the disease model or data, and $\delta_\epsilon$
to capture the nonlinear or second-order term. In quantitative
genetics, there is a similar approach in using the additive variance
$\sigma_a^2$ and dominance variance $\sigma_d^2$ 
(Fisher, 1918; Falconer and Mackay, 1996).
When these concepts from quantitative genetics are translated
to dichotomous traits, 
$\sigma_a^2= 2p(1-p)( p(f_2-f_1)+(1-p)(f_1-f_0))^2$, and
$\sigma_d^2= p^2(1-p)^2 ( f_2+f_0-2f_1)^2$
(see, e.g., Blackwelder and Elston, 1985).
Using Eq.(\ref{eq:delta-model}), we have
\begin{eqnarray}
\delta_p &=& \frac{f_2-K}{K(1-K)} p^2 + \frac{ f_1-K}{K(1-K)} p(1-p)
 \nonumber \\
&=& \frac{ pf_2+(1-p)f_1 - K}{K(1-K)} \cdot p
= \frac{ p(f_2-f_1)+q(f_1-f_0) }{K(1-K)} \cdot p(1-p).
\end{eqnarray}
In other words, the square of our first-order parameter
$\delta_p^2$ is proportional to the additive variance $\sigma_a^2$.
Similarly, using Eq.(\ref{eq:delta-model}) for $\delta_\epsilon$,
we have
\begin{eqnarray}
\delta_\epsilon &=& p^2(1-p)^2 \left( \frac{1}{K^2}-\frac{1}{(1-K)^2}\right)
(f_2f_0-f_1) +\left( \frac{p(1-p)}{1-K}\right)^2 (2f_1-f_2-f_0)
 \nonumber \\
&=& \frac{p^2(1-p)^2 (1-2K)}{K^2(1-K)^2} (f_2f_0-f_1^2)
+ \frac{p^2(1-p)^2}{(1-K)^2} (f_2+f_0 -2f_1).
\end{eqnarray}
When the disease prevalence $K$ is low, $\delta_\epsilon$
is roughly proportional to $f_2f_0-f_1^2$, as versus the
$f_2+f_0 -2f_1$ expression in $\sigma_d^2$. Actually, the
HWD coefficient in the control group, which is small, is proportional 
to the difference between $f_2f_0-f_1^2$ and $f_2+f_0 -2f_1$,
and two are approximately equal when $\lambda_1, \lambda_2$ are
small (Zheng et al., 2006b). 
Crudely, the square of $\delta_\epsilon$ can be said to be
proportional to the dominance variance $\sigma_d^2$.

The idea to use the test that is most powerful to the underlying
model sounds straightforward, but in reality the true disease model
is unknown. There have been attempts to infer the disease model 
by the HWD information.  For example, (Wittke-Thompson et al., 2005)  distinguishes 
HWD from different disease models and proposed its
use for data fitting  (note that the additive model defined in 
(Wittke-Thompson et al., 2005), $\lambda_2=2\lambda_1$, is different 
from that defined here, $f_2-f_1=f_1-f_0$ or $\lambda_2=2\lambda_1-1$).
In (Zheng and Ng, 2008), the signs of $\epsilon_1$ and $\epsilon_0$
are used for genetic model selection: (+,$-$) for recessive models,
($-$,+) for dominant models, and ($-$,$-$) for multiplicative and
additive models.  Since the amount of HWD in control 
group is usually much smaller than that in case group, 
the sign of $\delta_\epsilon = \epsilon_1- \epsilon_0 \approx \epsilon_1$
may serve the purpose in selecting recessive and dominant models.
All these previous works use HWD alone in genetic model selection, 
without considering a joint effect of HWD and allele frequency difference.

The result discussed in this paper shows that a joint consideration
of $\delta_\epsilon$ and $\delta_p$ could be more effective,
than a consideration of $\delta_\epsilon$ only, in selecting
disease model.
The following simple procedure  might be reasonable: 
first draw a line from origin to the data-determined
$(x,y)=(\delta_\epsilon,\delta_p)$ position, then calculate the angle formed
by this line and the $y$-axis. If the angle is smaller than
3$\pi$/32 (or 16.875$^\circ$),
the underlying model
is more likely to be multiplicative and CAT0.5 is the preferred test
to use. On the other hand, if the angle with respect to the $y$-axis
is larger, the underlying model is more likely to be recessive (if
it's in the first quadrant) or dominant (second quadrant), and MAX2
is preferred over CAT0.5. 
A caution on this procedure is that, unless the
sample size is very large and unless the true model is away from
the phase transition line, the disease model can still be
incorrectly inferred.

\begin{figure}[t]
  \begin{turn}{-90}
   \epsfig{file=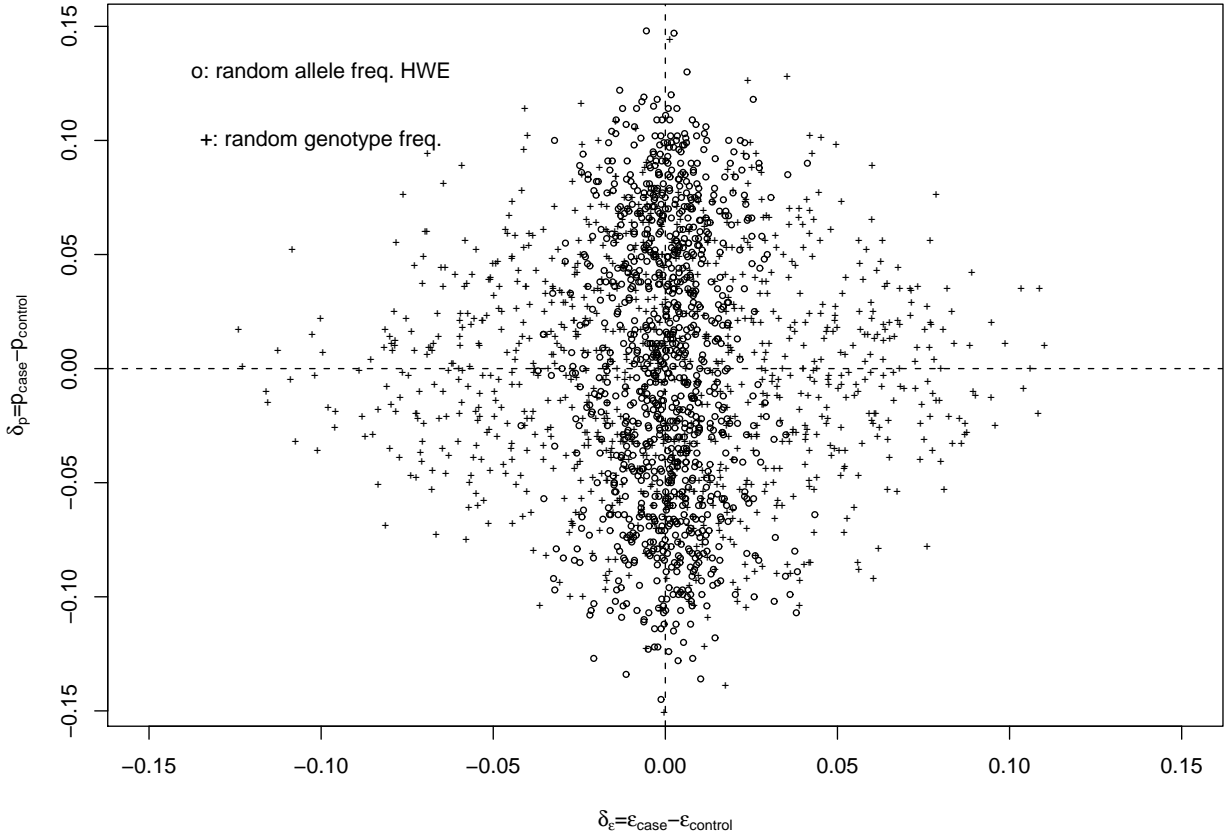, width=11cm}
  \end{turn}
\caption{
\label{fig5}
Distribution of datasets in two random sampling of disease models: 
(1) $p_{A, case}$ is randomly selected from (0.1, 0.9), and
$p_{A, control}$ from ($p_{A, case}$-0.1, $p_{A, case}$+0.1).
Genotype frequencies in cases and in controls follow the Hardy-Weinberg
equilibrium within the respective group. One replicate of
genotype counts is simulated which is used to calculated the
location in phase diagram by Eq.(\ref{eq:delta-data}). 
(2) Genotypes $aa, Aa, AA$ in case group  are randomly selected by the
interval probabilities: from 0 to $L_1$, from $L_1$ to $L_2$, and from $L_2$ to 1,
where $L_1$ and $L_2$ are sampled from (0.1,0.45) and (0.55, 0.9). The genotype
frequencies for control group are determined by the interval probabilities 
[0, $L_1'$, $L_2'$, 1] also, with $L_1'$ and $L_2'$ randomly sampled
from ($L_1$-0.1, $L_1$+0.1), ($L_2$-0.1, $L_2$+0.1).  
}
\end{figure}

Comparing the power of two tests is always tricky because 
the answer depends on what is known about the true model. 
It is very much in the spirit of Bayesian statistics
that the posterior probability distribution depends on the prior.
Any claim on discovering a more powerful test may contain
a caveat on how the alternative model is sampled.
Phase diagram discussed here provides a framework to visually 
inspect distributions on the $\delta_p$-$\delta_\epsilon$,
for the simulated models,

Fig.\ref{fig5} shows the distribution on the phase diagram
of two different ways of simulating genotype count tables.
The first is by randomly sampling two allele frequencies $p_{case}$
and $p_{control}$ that are close to each other (difference is less
than 0.1), then the genotype frequencies in the model are determined by
the HWE. One such model is used to generate
one replicate dataset and $\delta_\epsilon$, $\delta_p$ values are
determined from the replicate. The second way to generate a random
model is to randomly sample two sets of genotype frequencies, for
case and control group, that are close to each other (difference is 
less than 0.1), and that model is used to simulate one replicate
dataset. 

When we compare the empirical power of MAX2 and MAX3, MAX3 is
more powerful than MAX2 in the first simulation, whereas
MAX2 is slightly more powerful than MAX3 in the second simulation.
The phase diagram in Fig.\ref{fig5} easily solves the puzzle:
from Fig.\ref{fig5}, it can be seen that simulated datasets
by the first method centered around the $y$-axis, as genotype
frequencies in the model are obtained by HWE,
whereas those by the second method are more widespread in the $x$-axis. 
If we require a graphic showing of the phase diagram like
Fig.\ref{fig5} in any power comparison between two tests,
there would be less misunderstanding of seemingly conflicting empirical results.

\begin{figure}[t]
  \begin{turn}{-90}
   \epsfig{file=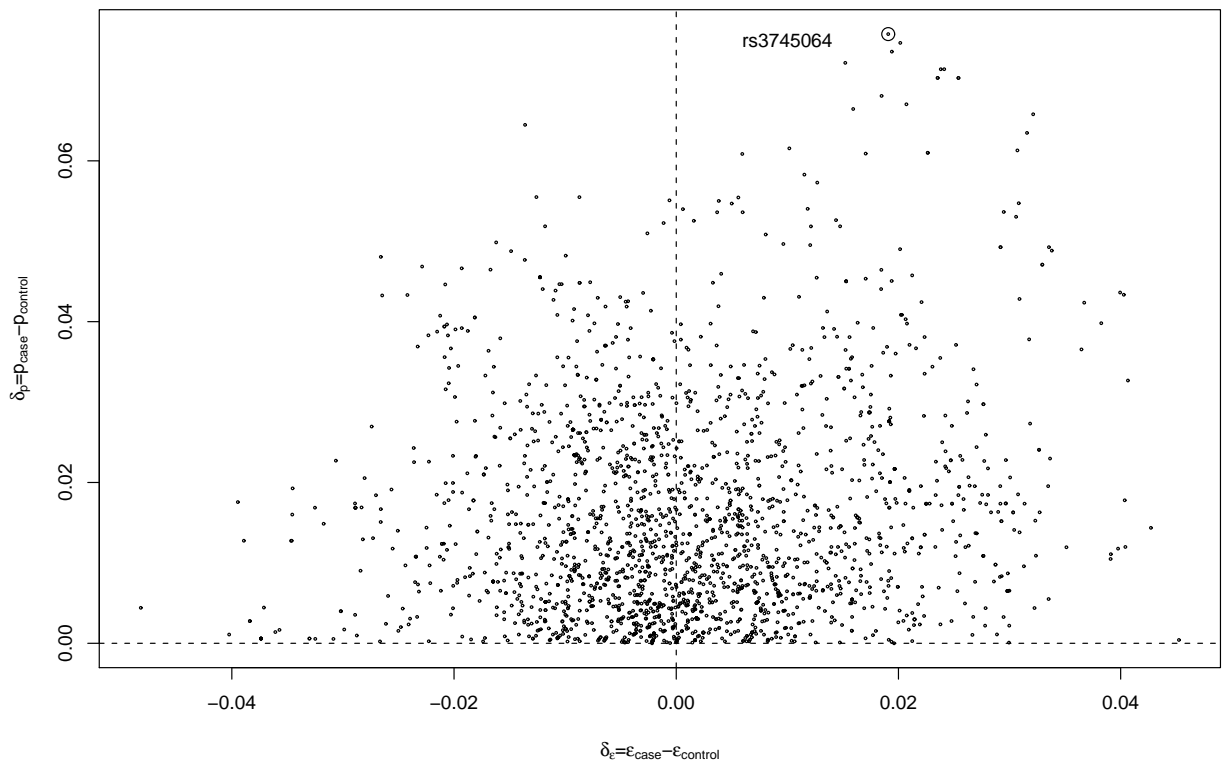, width=10cm}
  \end{turn}
\caption{
\label{fig6}
Distribution of 2147 SNPs on chromosome 18q of a case-control
dataset (Amos et al., 2007)   on the $\delta_\epsilon$-$\delta_p$
phase diagram. The SNP with the largest allele frequency
difference, rs3745064, is marked with a circle. Note that
rs3745064 is not on the $y$-axis, but inside the first
quadrant, indicating that the recessive model describes its
effect better than the multiplicative model.
}
\end{figure}

The phase-diagram can also be used to summarize a case-control
dataset with many SNPs. Fig..\ref{fig6} shows 2147 SNPs on chromosome
18q21 for 460 cases and 459 controls  used in Genetic Analysis 
Workshop 15 (Amos et al., 2007; Wilcox et al., 2007),
with $\hat{\delta}_p$ and $\hat{\delta}_\epsilon$
calculated by Eq.(\ref{eq:delta-data}).  These SNPs are filtered from
the original list of 2719 SNPs by requiring no more than 10 missing
typings and minor allele frequency to be larger than 0.05.
For some SNPs, alleles are switched to make sure $\delta_p$ is positive.
One striking visual impression of Fig.\ref{fig6} is that
the SNPs with the largest $\delta_p$ values are not located
on $y$-axis, but in the first quadrant, indicating that
recessive model better describes the effect of these SNPs
than the multiplicative model. 

In Fig.\ref{fig6}, the SNP rs3745064 at position 53.305Mb
exhibits the largest $\delta_p$ value (=0.076) 
(Kuo et al., 2007; Tapper et al., 2007). 
The distribution of case samples among the three genotypes
is 37, 180, 243, and that of control samples is 50, 223, 186,
leading to $p_1=0.724$, $p_0=0.648$, $\delta_1=0.0042$, $\delta_0=-0.015$.
For this SNP, we expect the largest test statistic value
to be CAT($x=0$) because it is located in the first quadrant
on the phase diagram.
Indeed, CAT($x=0$)=13.97, CAT($x=0.5$)=12.47, CAT($x=1$)=2.18, 
and MAX2 test statistic is more powerful than CAT($x=0.5$).

Using the phase diagram to examine the most recent whole-genome
association data shows that SNPs with the strongest association
signal tend to be multiplicative, and CAT0.5 test is more
powerful than MAX2 test in this situation. However, we should 
not exclude the possibility that it is due to a selection bias, as 
the top ranking genes were chosen by CAT0.5 test result. Also,
if the most significant SNPs exhibit larger allele frequency
differences, whereas their $\delta_\epsilon$ value is limited,
their positions in the phase diagram is expected to be closer to the $y$-axis.

Whether the result in Table 2(B) is due to selection bias or not,
our approach could be useful in analyzing whole-genome 
association data, as illustrated by Fig.\ref{fig6}. 
Applying MAX2 may change the rank order of
some SNPs that are near the top, and consequently change
the pool of SNPs to be studied further.  If MAX2 test does
improve the $p$-value over CAT0.5 for a SNP, one can check
whether the HWD mainly occurs in the case 
instead of the control group. The location of a SNP on the
phase diagram provides a quick filtering of SNPs where the 
inclusion of non-multiplicative models may improve the association 
signal, such as the example of rs3745064 on Fig.\ref{fig6}. 
Due to a high cost of typing extra samples in validating 
the associated SNPs for the second round, it is important 
to carry out careful analyses including the consideration 
of alternative disease models.

In conclusion, using the phase diagram is to
partition the violation of null hypothesis of equal genotype
frequency in case and control groups into two components,
one $\delta_p$ for allele frequency difference and another 
$\delta_\epsilon$ for difference in HWD coefficients. 
The relative magnitude of $\delta_p$ and $\delta_\epsilon$ 
determines which test, e.g., between CAT0.5 and MAX2, is more 
powerful. The phase diagram highlights the point that
no uniformly powerful test exists, and a test is only more
powerful regionally in the model space. 
We believe the use of phase diagram can aid the design 
of test when some knowledge of the mode of inheritance is available, 
as well as inferring the underlying mode of inheritance from the data.

\section*{ACKNOWLEDGEMENT}

WL acknowledgement the support from The Robert S. Boas Center for 
Genomics and Human Genetics at the Feinstein Institute for Medical
Research, and YY is supported by China NSF Grant (No. 10671189) and 
Chinese Academy of Science Grant (No. KJCX3-SYWS02). We would like
to thank the two reviewers for their comments and suggestions.

\section*{REFERENCES}

\vspace{0.09in}
\noindent
Amos CI (2007).
Successful design and conduct of genome-wide association studies,
Hum. Mol. Genet., 16, R220-R225.

\vspace{0.09in}
\noindent
Amos CI, Chen WV, Remmers E, Siminovitch KA, Seldin MF,
Criswell LA, Lee AT, John S, Shephard ND, Worthington J,
Cornelis F, Plenge RM, Begovich AB, Dyer TD, Kastner DL,
Gregersen PK (2007).
Data for Genetic Analysis Workshop (GAW) 15 Problem 2, genetic
causes of rheumatoid arthritis and associated traits,
BMC Proc.,  1(suppl 1), S3.

\vspace{0.09in}
\noindent
Balding DJ (2006).
A tutorial on statistical methods for population association studies.
Nat. Rev. Genet., 7,781-791.

\vspace{0.09in}
\noindent
Blackwelder WC, Elston RC (1985).
A comparison of sib-pair linkage tests for disease susceptibility loci,
Genet. Epid., 2, 85-97.

\vspace{0.09in}
\noindent
Devlin B \& Roeder K (1999).
Genomic control for association studies.
Biometrics, 55,997-1004.

\vspace{0.09in}
\noindent
Falconer DS, Mackay TFC (1996).
{\sl Introduction to Quantitative Genetics}, 4th edition,
(Benjamin Cummings).

\vspace{0.09in}
\noindent
Fisher RA (1918).
The correlation between relatives on the supposition of Mendelian inheritance,
Phil Tran Royal Soc Edinburgh,  52, 399-433.

\vspace{0.09in}
\noindent
Freidlin B, Zheng G, Li Z, Gastwirth JL (2002).
Trend tests for case-control studies of genetic markers,
power, sample size and robustness.
Hum. Heredity, 53,146-152.

\vspace{0.09in}
\noindent
Gibbs JW (1873).
Graphical methods in the thermodynamics of fluids,
{\sl Transactions of Connecticut Academy of Arts
and Science}, 2(11), 309-342;
A method of geometrical representation of the
thermodynamic properties of substances by
means of surfaces, {\sl ibid.}, 2(14), 382-404.

\vspace{0.09in}
\noindent
Gonz\'{a}lez JR, Carrasco JL, Dudbridge F, Armengol L,
Estivill X, Morento V (2008).
Maximizing association statistics over genetic models,
Genet. Epid., 32, 246-254.

\vspace{0.09in}
\noindent
Kuo TY, Lau W, Hu C, Zhang W (2007).
Association mapping of susceptibility loci for rheumatoid arthritis,
BMC Proc., 1(Suppl 1), S15.

\vspace{0.09in}
\noindent
Lewis CM (2002).
Genetic association studies, design, analysis and interpretation.
Brief. Bioinformatics, 3,146-153.

\vspace{0.09in}
\noindent
Li Q, Zheng G, Li Z, Yu K (2008).
Efficient approximation of p-value of the maximum of correlated tests, with 
applications to genome-wide association studies,
Ann. Hum. Genet., 72, 397-406.

\vspace{0.09in}
\noindent
Li W (2008).
Three lectures on case-control genetic association analysis.
Brief. Bioinformatics, 9,1-13.

\vspace{0.09in}
\noindent
Lifshitz EM \& Landau LD (1980).
{\sl Statistical Physics, Course of Theoretical Physics, Volume 5},
3rd edition (Butterworth-Heinemann).

\vspace{0.09in}
\noindent
Parkes M, Barrett JC, Prescott NJ, Tremelling M, Anderson CA, Fisher SA, Roberts RG, 
Nimmo ER, Cummings FR, Soars D (2007).
Sequence variants in the autophagy gene IRGM and multiple other replicating loci 
contribute to Crohn's disease susceptibility.
Nat. Genet., 39,830-832.

\vspace{0.09in}
\noindent
Sasieni PD (1997).
From genotypes to genes, doubling the sample size.
Biometrics, 53,1253-1261.

\vspace{0.09in}
\noindent
Scheet P \& Stephens M (2006).
A fast and flexible statistical model for large-scale population genotype data: 
applications to inferring missing genotypes and haplotypic phase.
Am. J. Hum. Genet., 78,629-644.

\vspace{0.09in}
\noindent
Slager SL \& Schaid DJ (2001).
Case-control studies of genetic markers, power and sample
size approximations for Armitage's test for trend.
Hum. Heredity, 52,149-153.

\vspace{0.09in}
\noindent
Suh YJ, Li W (2007).
Genotype-based case-control analysis, violation of Hardy-Weinberg equilibrium,
and phase diagram,
in Sankoff D, Wang L, Chin F (ed),
{\sl Proceedings of the 5th Asia-Pacific Bioinformatics Conference},
pp.185-194 (Imperial College Press).

\vspace{0.09in}
\noindent
Tapper W, Collins A, Morton NE (2007).
Mapping a gene for rheumatoid arthritis on chromosome 18q21,
BMC Proc.,  1(Suppl 1), S18.

\vspace{0.09in}
\noindent
Todd JA, Walker NM, Cooper JD, Smyth DJ, Downes K, Plagnol V, Bailey R, Nejentsev S, Field SF, 
Payne F (2007).
Robust associations of four new chromosome regions from genome-wide analyses of type 1 diabetes.
Nat. Genet., 39,857-864.

\vspace{0.09in}
\noindent
Tokuhiro S, Yamada R,  Chang X,  Suzuki A,  Kochi Y,  Sawada T, Suzuki M,
Nagasaki M,  Ohtsuki M,  Ono M,  Furukawa H,  Nagashima M,  Yoshino S, Mabuchi A,
 Sekine A,  Saito S,  Takahashi A,  Tsunoda T,  Nakamura Y,   Yamamoto K (2003).
An intronic SNP in a RUNX1 binding site of SLC22A4, encoding
an organic cation transporter, is associated with rheumatoid arthritis.
Nat. Genet., 35,341-348.

\vspace{0.09in}
\noindent
Ury HK (1976).
A comparison of four procedures for multiple comparisons among
means (pairwise contrasts) for arbitrary sample sizes.
Technometrics, 18,89-97.

\vspace{0.09in}
\noindent
Weir BS (1996).
{\sl Genetic Analysis II} (Sinauer Associates, Sunderland, MA).

\vspace{0.09in}
\noindent
Wellcome Trust Case Control Consortium (2007).
Genome-wide association study of 14,000 cases of seven common diseases and 
3,000 shared controls,
Nature, 447,661-678.

\vspace{0.09in}
\noindent
Wellcome Trust Case Control Consortium \&  Australo-Anglo-American Spondylitis Consortium 
(TASC) (2007).
Association scan of 14,500 nonsynonymous SNPs in four diseases identifies autoimmunity 
variants, Nat. Genet., 39,1329-1337.

\vspace{0.09in}
\noindent
Wilcox MA, Li Z, Tapper W (2007).
Genetic association with rheumatoid arthritis - Genetic Analysis Workshop 15: 
summary of contributions from Group 2,
Genet. Epid,  31(S1), S12-S21.

\vspace{0.09in}
\noindent
Wittke-Thompson JK, Pluzhnikov A, Cox NJ (2005).
Rational inferences about departures from Hardy-Weinberg
equilibrium.
Am. J. Hum. Genet., 76,967-986.

\vspace{0.09in}
\noindent
Yamada R, T. Tanaka, M. Unoki, T. Nagai, T. Sawada, Y. Ohnishi, T. Tsunoda,
M.  Yukioka, A. Maeda, K. Suzuki, H. Tateishi, T. Ochi, Y. Nakamura, K. Yamamoto
(2001).
Association between a single-nucleotide polymorphism in the promoter
of the human interleukin-3 gene and rheumatoid arthritis in Japanese
patients, and maximum-likelihood estimation of combinatorial effect
that two genetic loci have on susceptibility to the disease.
Am. J. Hum. Genet., 68,674-685.

\vspace{0.09in}
\noindent
Zeggini E, Weedon MN, Lindgren CM, Frayling TM, Elliott KS, Lango H, 
Timpson NK, Perry JRB, Rayner NW, Freathy RM, Barrett JC, Shields B, Morris AP,
Ellard S, Groves CJ, Harries LW, Marchini JL, Owen KR, Knight B, Cardon LR, 
Walker M, Hitman GA, Morris AD, Doney ASF, The Wellcome Trust Case Control 
Consortium (WTCCC). McCarthy MI, Hattersley AT (2007),
Replication of genome-wide association signals in UK samples reveals risk 
loci for type 2 diabetes.
Science, 316,1336-1341.
	
\vspace{0.09in}
\noindent
Zheng G, Freidlin B, Gastwirth JL (2006a).
Comparison of robust tests for genetic association using
case-control studies.
in
{\sl Optimality, The Second Erich L. Lehmann Symposium --
IMS Lecture Notes Vol.49}, ed. Rojo J, pp.253-265
(Institute of Mathematical Statistics).

\vspace{0.09in}
\noindent
Zheng G, Freidlin B, Gastwirth JL (2006b).
Robust genomic control for association studies,
Am. J. Hum. Genet.,  78, 350-356.

\vspace{0.09in}
\noindent
Zheng G, Meyer G, Li W, Yang Y (2008).
Comparison of two-phase analyses for case-control association studies,
Stat. Med., to appear.

\vspace{0.09in}
\noindent
Zheng G, Ng HKT (2008).
Genetic model selection in two-phase analysis for
case-control association studies.
Biostat., 9, 391-399.

\begin{table}
\begin{center}
(A) genotype count table \\
\begin{tabular}{ccccc}
\hline
 & sample size &  aa & aA & AA   \\
\cline{3-5} 
case(1) & $N_1$ &  $N_{10}$ & $N_{11}$ & $N_{12}$ \\
control(0) & $N_0$ &  $N_{00}$ & $N_{01}$ & $N_{02}$ \\
\hline
\end{tabular}
\vspace{0.5in}

(B) same genotype count table parameterized by $p_1, p_0, \epsilon_1, \epsilon_0$\\
\begin{tabular}{cccc}
\hline
 &  aa & aA & AA \\
\cline{2-4}
case(1) & 
 $N_1 [ (1-p_1)^2 + \epsilon_1]$ &
 $N_1 [2 (1-p_1) p_1  -2 \epsilon_1]$ &
 $N_1 [ p_1^2  + \epsilon_1]$  \\
control(0) &
 $N_0 [ (1- p_0)^2 + \epsilon_0]$ &
 $N_0 [2 (1- p_0) p_0  -2 \epsilon_0]$ &
 $N_0 [ p_0^2  + \epsilon_0]$ \\
\hline
\end{tabular}
\caption{
Genotype count table (A) and its parameterization using
$p_1, p_0, \epsilon_1, \epsilon_0$ (B).
}
\end{center}
\end{table}

\begin{table}
\begin{center}
(A) genotype count table \\
\begin{tabular}{cccccc}
\hline
disease & gene & SNP & chromosome & case(aa/aA/AA) & control (aa/aA/AA) \\
\hline
AS$^a$ & ARTS1 & rs27044 &  5   & 793/553/119   & 395/432/94 \\
AS &LNPEP &rs2303138 &5     &738/176/8      &1269/193/4 \\
T1D$^b$ &C12orf30 &rs17696736&12&1984/3115/1168 &1545/2891/1373 \\
T1D &ERBB3 &rs2292239 &12   &2592/2805/801  &1956/2816/946 \\
T1D &KIAA0350 &rs12708716&16&2652/2857/779  &2834/2429/569 \\
T1D &PTPN2 &rs2542151 &18   &4219/1635/182  &3628/1889/220 \\
T1D &PTPN22 &rs2476601 &1   &4674/998/55    &3754/1580/178 \\
CD$^c$ &IRGM &rs13361189 &5     &2061/705/59    &8907/1476/54 \\
CD &NOD2 &rs17221417 &16    &1505/1175/256  &747/754/245 \\
CD &IL23R &rs11805303 &1    &1385/1236/313  &655/815/276 \\
T2D$^d$ &FTO &rs8050136 &16     &1063/1407/464  &550/987/378 \\
\hline
\end{tabular}

(B) position on the phase diagram \\
\begin{tabular}{cccccccccc}
\hline
disease/SNP &  $p_1$ & $p_0$ & $\delta_p$ &
 $\epsilon_1$ & $\epsilon_0$ & $\delta_\epsilon$ & $\theta$ & $\theta-90^\circ$ & CAT(0.5/0/1)  \\ 
\hline
AS/rs27044    & .34  & .27 & .067 & $-$.011 &  .0083 & $-$.020 & 106.4 & 16.4 &23.9/3.0/{\bf 28.6}\\
AS/rs2303138  & .069 & .10 & $-$.036 & $-$.0020 & $-$.0022 & $2 \times 10^{-4}$ & $-$89.7 & 0.3 & 19.4/4.0/17.9 \\
T1D/rs17696736& .49  & .43 & .050 & $9 \times 10^{-4}$ & $-$.0028 & .0037 & 85.8 & -4.2 & 61.5/45.3/37.3 \\
T1D/rs2292239 & .41  & .36 & .056 & $-$.004 & .0028 & $-$.0069 & 97.0 & 7.0 & 79.3/31.2/73.0 \\
T1D/rs12708716& .31  & .35 & $-$.045 & .004 & $ 6 \times 10^{-4}$ & .0034 & $-$85.7 & 4.3 & 55.4/21.2/50.3 \\ 
T1D/rs2542151 & .20  & .17 & .037 & $-$.0029 & .0027 & $-$.0056 & 98.5 & 8.5 & 54.7/5.6/{\bf 58.7} \\
T1D/rs2476601 & .18  & .097& .079 & .0015 & $ 2 \times 10^{-4}$ & .0012 & 89.1 & -0.9 & 292.6/71.2/273.2 \\
CD/rs13361189 & .076 & .15 & $-$.070 & $-6 \times 10^{-4}$& $- 3 \times 10^{-4}$ & $-3 \times 10^{-4}$& $-$90.2 & -0.2 & 261.5/65.0/238.4 \\
CD/rs17221417 & .36 & .29 & .069 & .013 & .0047 & .0088 & 82.8 & -7.2 & 46.5/32.3/31.5 \\
CD/rs11805303 & .39 & .32 & .074 & .0048 & .0060 & $-$.0012 & 90.9 & 0.9 & 51.7/26.3/41.8 \\ 
T2D/rs8050136 & .46 & .40 & .057 & $-$.0097 & $-2 \times 10^{-4}$ & $-$.0095 & 99.5 & 9.5 & 31.4/12.4/29.4 \\
\hline
\end{tabular}
\caption{
(A) Genotype counts of SNPs that show strong (and validated) association signal with several
common diseases. a. ankylosing spondylitis (Wellcome Trust Case Control Consortium \& 
 Australo-Anglo-American Spondylitis Consortium, 2007). 
b. type 1 diabetes (Todd et al., 2007).
c. Crohn's disease (Parkes et al., 2007).
d. type 2 diabetes (Zeggini et al., 2007).
Other information includes gene and SNP name, as well as the chromosome number on
which the SNP is located.
(B) For the SNP listed in (A), minor allele frequency in case ($p_1$) and control ($p_0$) group, their
difference ($\delta_p = p_1-p_0$), HWD
coefficients in case ($\epsilon_1$) and control ($\epsilon_0$) group,
and their difference ($\delta_\epsilon= \epsilon_1-\epsilon_0$),
angle $\theta = tan^{-1} (\delta_p/\delta_\epsilon)$ in the phase diagram,
angle with respect to the $y$-axis ($\theta-90^\circ$ if $\theta >0$, or
$\theta + 90^\circ$ if $\theta <0$), and
test statistics for CAT0.5, CAT0, CAT1.
}
\end{center}
\end{table}

\newpage

%######################### fig1

%######################### fig2

%######################### fig3

%######################### fig4

%######################### fig5 (new)

%######################### fig6 (new)

\end{document}